# Decaying shock studies of phase transitions in MgO-SiO$_2$ systems: implications for the Super-Earths' interiors.

R. M. Bolis[1], G. Morard[2], T. Vinci[1], A. Ravasio[1], E. Bambrink[1], M. Guarguaglini[1], M. Koenig[1,3], R. Musella[4], F. Remus[4,5], J. Bouchet[5], N. Ozaki[6,7], K. Miyanishi[7], T. Sekine[8], Y. Sakawa[9], T. Sano[9], R. Kodama[3,6,7], F. Guyot[2], A. Benuzzi-Mounaix[1],

[1] LULI - CNRS, Ecole Polytechnique, CEA : Université Paris-Saclay; UPMC Univ. Paris 06; Sorbonne Universités - F-91128 Palaiseau cedex, France.

[2] Institut de Minéralogie, de Physique des Matériaux et de Cosmochimie, UMR CNRS 7590, Sorbonne Universités - Université Pierre et Marie Curie, CNRS, Muséum National d'Histoire Naturelle, IRD, 4 Place Jussieu, 75005 Paris, France.

[3] Institute for Academic Initiatives, Osaka University, Suita, Osaka 565-0871, Japan .

[4] LUTH, Observatoire de Paris, CNRS, Université Paris Diderot, 92195 Meudon, France.

[5] CEA, DAM, DIF, 91297, Arpajon, France

[6] Graduate School of Engineering, Osaka University, Suita, Osaka 565-0871, Japan

[7] Photon Pioneers Center, Osaka University, Suita, Osaka 565-0871, Japan

[8] Department of Earth and Planetary Systems Science, Hiroshima University, Higashi-hiroshima 739-8526, Japan

[9] Institute of Laser Engineering, Osaka University, 2-6 Yamada-oka, Suita, Osaka 565-0871, Japan

Corresponding author: Riccardo Maria Bolis (riccardo.bolis@polytechnique.edu)

**Key Points:**

- Magnesium oxides phase diagram investigated along the Hugoniots between 0.2 and 1.2 TPa with decaying shocks and optical diagnostics
- New MgO experimental melting point proposed at 0.47 TPa and 9860 K
- No evidences of phase transition have been found for MgSiO$_3$ and Mg$_2$SiO$_4$ between 0.12-0.5 TPa and 0.2-0.85 TPa respectively.

Abstract

We report an experimental study of the phase diagrams of periclase (MgO), enstatite (MgSiO$_3$) and forsterite (Mg$_2$SiO$_4$) at high pressures. We investigated with laser driven decaying shocks the pressure/temperature curves of MgO, MgSiO$_3$ and Mg$_2$SiO$_4$ between 0.2-1.2 TPa, 0.12-0.5 TPa and 0.2-0.85 TPa respectively. A melting signature has been observed in MgO at 0.47 TPa and 9860 K, while no phase changes were observed neither in MgSiO3 nor in Mg2SiO4. An increasing of reflectivity of MgO, MgSiO$_3$ and Mg$_2$SiO$_4$ liquids have been detected at 0.55 TPa -12 760 K, 0.15 TPa - 7540 K, 0.2 TPa - 5800 K, respectively. In contrast to SiO$_2$, melting and metallization of these compounds do not coincide implying the presence of poor electrically conducting liquids close to the melting lines. This has important implications for the generation of dynamos in Super-earths mantles.

1 Introduction

MgO, MgSiO$_3$ and Mg$_2$SiO$_4$ are among the most relevant magnesian end members components of the Earth mantle and generally of telluric planets. MgO is issued from the dissociation of (Mg,Fe)$_2$SiO$_4$ into (Mg,Fe)SiO$_3$ and (Mg,Fe)O at 21 GPa following the geotherm and represents the Mg-rich end member of (Mg,Fe)O ferropericlase, the second most abundant mineral in planetary mantles. For this reason, a detailed knowledge of the phase diagram of these magnesian end-members components is essential to properly model interior structures and dynamics of telluric planets. As Mg, Si and O dominate by far the chemical composition of their mantles, the study of the magnesian end-members allows to model the interior structures at the first order, avoiding the experimental and theoretical complexities induced by more minor elements (such as Ca, Al or Fe). While Diamond Anvil Cell (DAC) measurements have substantially contributed to the knowledge of the upper regions of the mantles, our comprehension of the processes governing the deeper mantles is limited mainly due to the inadequate characterization of magnesium oxide and silicates properties at high pressures. Indeed while experimental and theoretical data are still scarce even at the moderate pressure of Earth's core-mantle boundary (CMB) 130 GPa, thermal profiles of Super Earth's interiors are controlled by the melting lines of these compounds at even higher pressures [*Stixrude*, 2014]. For example, the CMB in GJ876d, a super Earth with 7.5 Earth masses, is expected to be seated at ~1000 GPa [*Valencia et al, 2006*]. Moreover thermal and transport properties of these compounds at high pressures are also necessary to comprehend planetary evolution and macroscopic characteristics, such as the generation of sustained magnetic fields. Indeed, the increasing electrical conductivity of liquid silicates under extreme conditions and their transition to a metallic behavior would promote part of the mantle to a potential magnetic field generator, similarly to the Fe-rich core [e.g. *McWilliams et al.*, 2012; *Millot et al.*, 2015]. It is therefore a strong requirement to study these mineral phase diagrams over a wider pressure and temperature range. To this extent, dynamic compression schemes represent a valid alternative to achieve high pressures/temperatures conditions. Various theoretical and few experimental studies have lately been carried on dynamically compressed MgO, MgSiO$_3$ and Mg$_2$SiO$_4$ phases but some controversies are not solved yet.

Calculations [*Cebulla and Redmer*, 2014; *Boates and Bonev*, 2013] and experiments [*Zerr and Boehler*, 1993; *Coppari et al.*, 2013] agree that MgO phase diagram is composed by two solid crystalline phases (with B1 and B2 structures) and a liquid phase. However, at high pressure, phase boundaries in the P-T diagram remain uncertain. Recent decaying [*McWilliams et al.*, 2012] and steady [*Miyanishi et al.*, 2015; *Root et al.*, 2015] shock experiments claim to have observed signatures of melting and of the B1-B2 transitions, but at discordant P-T conditions. In addition, it has been observed that along the Hugoniot, MgO melts directly into a metallic liquid [*McWilliams et al.*, 2012], suggesting that magnesium oxide melt could play an important role in magnetic field generation in terrestrial planets.

The high-pressure phase diagram of $MgSiO_3$ has been probed along the Hugoniot. Below the melting line, calculations predict two solid phases with perovskite and post-perovskite structures, although not experimentally evidenced in dynamic compression experiments. In the $MgSiO_3$ melt region, a peculiar liquid-liquid phase transition has been observed in decaying shocks experiments, from both glass and single-crystal starting materials [*Spaulding et al.*, 2012]. However either this transition is not predicted [*Militzer*, 2013] or its nature is contested [*Boates and Bonev*, 2013]. This controversy together with the strong implications that a liquid-liquid transition would have for planetary interiors encourage a deeper investigation.

The $Mg_2SiO_4$ phase diagram has been explored with shock compression up to 200 GPa [*Mosenfelder et al.*, 2007], but no data are yet reported at higher pressures. A key question concerns $Mg_2SiO_4$ dissociation along the Hugoniot and this is the reason why it would be important to evidence possible signatures of MgO or $MgSiO_3$ in the $Mg_2SiO_4$ shock studies.

Decaying shocks are an efficient method to study phase transitions in transparent materials. A reflecting and not sustained shock is launched into an initially transparent material and its propagation is monitored measuring the time evolution of thermal emission, shock velocity and optical reflectivity at the shock front. As the shock is not sustained, the pressure, temperature and shock velocity gradually decrease in time, following the Hugoniot curve. Phase transition may exhibit strong variations in the temperature and/or in the shock speed due to release of latent heat or to strong volume change respectively. To detect these variations with optical diagnostics, the release of latent heat or the volume change must be large enough to produce observable imprints on the measured profiles. Moreover, the kinetics of the transition must be fast enough to occur before the shock front has propagated over distances exceeding the optical depth.

We conducted experimental campaigns exploring MgO, $MgSiO_3$ and $Mg_2SiO_4$ with decaying shocks. Our results claim a reinterpretation of the behavior of these magnesian endmember compounds along the shock Hugoniot up to now proposed.

**2 Materials and Methods**

We performed experimental campaigns on the LULI2000 laser facility at the Laboratoire Pour l'Utilisation des Lasers Intenses (LULI) of the Ecole Polytechnique in Palaiseau, France and on the GEKKO laser of the Institute of Laser Engineering (ILE) at the Osaka University, Japan.

In the LULI experiment we used two frequency doubled laser beams (wavelength = 0.53 µm) with a 1.2-1.5 ns pulse duration delivering up to 800 J on target. The focal spot was spatially smoothed with Hybrid Phase Plate resulting in a flat intensity profile of 600µm diameter. Similarly, at ILE, we used 6-9 beams of 2.5 ns at 0.351 µm focused with a Kinoform Phase Plate to 600 µm focal spot diameter. The total energy was ≤ 900 J. The resulting laser intensities achieved in these experiments were varied in the range $1-6 \times 10^{13}$ W/cm$^2$.

The targets were made of multilayer pushers glued on 350 µm thick samples: MgO single-crystal, $Mg_2SiO_4$ single-crystal and $MgSiO_3$ glass (See the Supporting information for details). The thickness of the glue was measured ≈ 1 µm. An anti-reflecting coating for 532 nm and 1064 nm was applied on the rear side of samples.

The choice of pushers is critical in decaying shocks experiments. Indeed the occurrence of any preheating issued from X-ray emission from coronal plasma and of reverberation waves can affect shock velocity and thermal emission signals, whose interpretation then becomes extremely complicated. To reduce the problem of preheating, we used a low-Z plastic polymer ablator followed by a high-Z X-ray absorbing layer. Reverberation waves are produced at interfaces with a high impedance mismatch (e.g. CH-Au). Problems rise when the reverberation catches the measured shock in the sample material, producing a jump in velocity and temperature signals that might be interpreted as a phase transition. To avoid this, the pusher must be optimized in such a way that this jump takes place inside the pusher itself and not in the sample material. To check these issues we tested three different types of pusher (Figure 1a):

1. 10µm CH / 2µm Au / 50µm Al;
2. 50µm CH / 15µm Fe;
3. 75 µm CH / 3 µm Au / 25 µm Ti.

A study conducted on $SiO_2$ as a reference material, allowed us to establish that pushers 2 and 3 are suitable for our measurements, while pusher 1 is not appropriate for decaying shock experiment on thick silicate samples [*Supporting information*].

The two main diagnostics employed were VISARs (Velocity Interferometer System for Any Reflector) interferometers at two wavelengths (532 nm and 1064 nm) and a Streaked Optical Pyrometer (SOP). VISARs allow to precisely measure the shock velocity and reflectivity as function of time with sensitivities described in *Supporting Information*.

In order to extract pressures (P) from shock velocity (Us) measurements we used interpolation of experimental data found in literature and/or theoretical models of equation of state (EoS): for MgO, the data of [*Root et al.*, 2015] and the Mie-Grüneisen-Debye model of [*McWilliams et al*, 2012]; for $MgSiO_3$, laser data from Spaulding et al. [2012] and the QEOS model; for $Mg_2SiO_4$ the QEOS model [*Supporting information; More et al., 1988*]. We verified that the relation P-Us is weakly dependent on the EoS model. This weak sensitivity has been taken into account in the error bar. SOP diagnostic coupled to reflectivity data allows us to calculate temperature using the gray body emission hypothesis. We calibrated our diagnostic with a temperature standard and validated the calibration with decaying shocks in the reference material ($SiO_2$) [*Supporting information*].

# 3 Results

## 3.1 MgO

MgO shock velocity (red curve) and thermal emission (blue curve) are shown in figure 1b. While the shock velocity decays monotonically, the thermal emission profile exhibits a clear "bump" (at ~11 ns for the shot shown in figure 1b). The thermodynamic conditions associated to this incident cannot be directly inferred form the optical data since it occurs in a regime where MgO is opaque and shock velocity cannot be measured (disappearance of the fringes in figure1b). Therefore, shock velocities to be associated with thermal emissivity in this phase diagram region, are obtained using extrapolation methods, as presented in the Supplementary Material of [*McWilliams et al.*, 2012]. The corresponding P-T curves are shown in figure 2a.

It is important to underline that we are able to reproduce the bump at the same T-P conditions with shocks generated with different laser pulses. As examples, in figure 2a we show shots 91, 44 and 83 that were performed with a 1.2 ns laser pulse of 265, 448.6 and 726.7 J respectively. This implies that the bump is a signature of a phase transition and it is not produced by reverberating waves. Indeed, in the latter case, for different laser conditions, the bump would be expected to occur at different pressures depending on the dynamic of the shock (velocity and decaying rate). Moreover, our data are coherent with another criterion which allows to discriminate between reverberations and phase transition looking at the curvature of the thermal emission signal [*Millot*, 2016].

Since recent calculations [*Cebulla and Redmer*, 2014; *Miyanishi et al.*, 2015; *Root et al.*, 2015] agree on the prediction that the Hugoniot curve crosses melting and B1-B2 lines, our interpretation is that the bump is a signature of the melting of the B2 phase. This interpretation is based on different arguments. First, recent ab initio simulations of the Hugoniot predict an important discontinuity along the Hugoniot at melting of the B2 with $\Delta P \approx 120$ GPa against $\Delta P \approx 40$ GPa at the B1/B2 boundary [*Miyanishi et al.*, 2015; *Root et al.*, 2015]. As a shock can follow a thermodynamic path out of equilibrium (which is our case, as explained in the following point), we do not expect to reproduce exactly the $\Delta P$ calculated at the equilibrium but we do expect that the signature would be more pronounced for melting than for the B1/B2 phase transition. Second, we observed that changing the laser intensity, and thus the decay dynamics, the bump amplitude changed (figure S7), supporting the idea of a non-equilibrium phenomenon. Moreover the shape of the bump is typical of superheating occurring when the temperature rises faster than the rate of rearrangement of atoms required for melting [*Luo and Ahrens*, 2004]. To better understand this interpretation, it is worth noting that a decaying shock experiment is equivalent to a series of dynamic compressions of succeeding weaker shock strengths. This means that during the shock propagation different layers of the sample are compressed and heated from the same initial cold state. Similar features to the one we measured were observed and attributed to superheating in previous laser and gas-gun experiments on alpha-quartz, stishovite and fused silica [*Lyzenga et al.*, 1983; *Hicks et al.*, 2006; *Millot et al.,* 2015]. Third, to investigate the possibility to detect the B1/B2 transition, we performed ab initio calculations estimating the MgO optical depth. The calculated optical depth is larger than 10 μm, at ≈ 325 GPa (i.e. around the B1/B2 expected transition) whereas it is less than 1 μm for conditions corresponding to the melting point

(470 GPa) detected here. In the first case, the thermal emission is thus not only coming from the shock front but from an extended inhomogeneous region behind it at off-Hugoniot conditions [see *Supporting information*]. The emissivity (i.e. temperature) data in these thermodynamic states are not suitable for a reliable phase transition detection along the Hugoniot.

We therefore interpret the only feature observed in our experiments as the melting signature of the B2 phase. Nevertheless, a different scenario where the Hugoniot would cross liquid/ B1 boundary, skipping the B2 phase, cannot be excluded. Direct structural investigations by using in-situ x-ray diffraction during the laser-driven compression experiments will be mandatory to resolve this issue.

Our measurements on MgO highlight a melting point at 0.47 ± 0.04 TPa and 9860 ± 810 K (0.85 ± 0.07 eV). This result is not far from recent ab-initio calculations [*Root et al.*, 2015; *Miyanishi et al*, 2015], but partially in contrast with previous experimental results [*McWilliams et al.*, 2012], where melting was determined at 0.6 TPa along the Hugoniot, as inferred from a slope change in the P-T curve (pointed by the black arrow in figure 2a). Our result has important implications if we consider the reflectivity data (figure 2b). Indeed, reflectivities start to smoothly increase at ~0.55 TPa, i.e. at higher pressures than the measured melting point (at 0.47 TPa and 9860 K), suggesting that melting is not directly associated with metallisation. We also observe that between 0.55 and 1 TPa, the reflectivity at 532 nm ($R(2\omega)$) is similar to the reflectivity at 1064 nm ($R(\omega)$), while for pressures higher than 1 TPa $R(\omega) > R(2\omega)$, saturating at ~ 20% and ~13% respectively. This behavior suggests a transition between a "semiconductor" to a metal. We can therefore apply Drude semiconductor model to obtain conductivity values from reflectivity measurements. This approach applied to MgO P-T states along the Hugoniot gives a conductivity around $10^4$ S/m for 1% reflectivity [*McWilliams et al.*, 2012]. This means that along the Hugoniot between 0.47 and ~ 0.55 TPa, MgO liquid has a conductivity lower that $10^4$ S/m. In particular of ~$10^3$ S/m at 0.47 TPa using extrapolation of this model. These results are confirmed by more sophisticated ab-initio calculations, which predict similar values, of ~ 7 x $10^4$ S/m for a 2% reflectivity. Such low conductivities are close to the minimum value for the dynamo mechanism to be operating in a molten layer at the bottom of a silicate mantle [*Ziegler and Stegman*, 2013]. This result is an important input for evaluating liquid MgO contribution to the generation of a planetary magnetic field.

3.2 MgSiO$_3$

MgSiO$_3$ thermal emission at the shock front decreases exponentially with shock velocities exhibiting the typical behavior of a shock that compress a material in the same phase during the entire propagation. Indeed, for MgSiO$_3$ glass, we have not observed any bumps, in contrast to a recent decaying shock experiment where a liquid-liquid phase transition was detected [*Spaulding et al.*, 2012]. MgSiO$_3$ glass pressure/temperature curves along the Hugoniot obtained at LULI and at GEKKO are shown in figure 3a. Since our shocks start their propagations in the melt region of the phase diagram [*Shen et al.*, 1995; *Belonoshko et al.*, 2005; *De Koker and Stixrude*, 2009], the only possible interpretation from our data is that the shocks propagated through one single liquid phase. It is worth doing some considerations that strengthen our results. First, because of the study of the hydrodynamics of our pusher, we expect that in our targets no hydrodynamic effects could have affected shock decays [*Supporting information*]. Moreover our results have been reproduced at two different facilities (LULI2000 and GEKKO). Furthermore below 0.4 TPa our Hugoniot data are in agreement with Spaulding et al. [2012]. The only disagreement concerns the feature at 0.4 TPa that has been proposed as a liquid-liquid phase transition. Such first order phase transition is quite rare (occurring in liquid phosphorous [*Katayama et al.,* 2000]), and for $MgSiO_3$ it has not been reproduced by molecular dynamics calculations [*Militzer,* 2013]. Our dataset does not rule out changes in Si coordination number by oxygen under extreme pressure or other structural transitions in the liquid state, but it implies that their effects on decaying shock would be too weak to be observed with optical diagnostics.

The reflectivity has been observed to increase above 0.15 TPa as pointed by the arrow in figure 3a. No signatures of transitions have been detected and since we can reasonably affirm that the shock initially melts the material (it starts its propagation at 0.5 TPa-22000 K), the absence of transition signature until the detection limit suggests to set an upper limit for melting at 6300 ± 690 K at 0.12 TPa. In addition the results show that melting and reflectivity increases are not coincident. This suggests the presence of a poorly conducting liquid also in the $MgSiO_3$ phase diagram just above the melting line.

### 3.3 $Mg_2SiO_4$

The high-pressure phase diagram of $Mg_2SiO_4$ is poorly know beyond 200 GPa. Open questions remain even at lower pressures on whether chemical and structural changes observed in static compression experiments persist under dynamic loading. In particular, it is not known whether forsterite breaks down into $MgO$-$MgSiO_3$ - $MgO$-$SiO_2$ systems or melts into pure $Mg_2SiO_4$ [*Mosenfelder et al.*, 2007; *De Koker et al.*, 2008]. Hence, for $Mg_2SiO_4$, we looked both for any signatures of phase transitions, such as dissociation or melting, along the Hugoniot as well as for any clues of the occurrence of MgO at the crossing between the Hugoniot and MgO phase boundaries. We obtained the pressure/temperature curves shown in figure 3b, where the shock pressure decays monotonically between 850 and 200 GPa. In this region neither velocity nor emissivity profiles exhibit signatures of phase transitions. In addition, no features have been observed at the crossing between the Hugoniot and MgO phase boundaries. In particular the data do not show the melting signature detected at 0.47TPa and 9860 K in MgO samples. These results either exclude the presence of MgO (i.e. $Mg_2SiO_4$ remains in a single undissociated phase in the pressure range explored, potentially due to the slow dissociation kinetic) or suggest that, because of an eutectic behavior in the $MgO$-$SiO_2$ or $MgO$-$MgSiO_3$, MgO concentration is not high enough to generate a detectable signature with decaying shocks.

The $Mg_2SiO_4$ reflectivity increase has been observed starting at 0.2 TPa. No melting signatures have been detected down to 0.2 TPa and 6300 ± 680 K, the detection limit of our diagnostics. This observation is in agreement with previous measurements [*Luo et al.*, 2004] which set forsterite melting between 150 GPa and 170 GPa along the Hugoniot. Our data allow to uderline that, as for the cases of MgO and $MgSiO_3$, melting does coincide with the onset of reflectivi-ty, suggesting the existence of a poorly conductive liquid in the $Mg_2SiO_4$ high pressure phase diagram in the vicinity of the melting line.

# 4 Conclusions

In this study, we have obtained pressure/temperature measurements along the Hugoniot of different compounds in the $MgO-SiO_2$ chemical system.

We propose an upper limit of the melting point of MgO at 0.47 ± 0.04 TPa and 9860 ± 810 K (0.85 ± 0.07 eV). This is lower than several estimates of melting in this pressure range but still implies that, for the quasi totality of the planetary mantles of the super Earths, MgO will remain in the solid state [*Stixrude,* 2014]. As iron should be present in planetary mantles, future experiments on FeO and (Mg,Fe)O solid solutions should however be carried out in order to better constrain conditions of partial melting in this system. Moreover, molten MgO could play an important role in the case of deep magma oceans and/or in response to very large impacts, as well as in the putative cores of icy giant and gas giant planets (e.g. for tidal dissipation, see [*Remus et al.,* 2012]). The scenario proposed here with a lower pressure/temperature melting point brings a new important input for modeling studies of these phenomena.

We have also shown that no detectable phase transition could be observed neither in $MgSiO_3$ nor in $Mg_2SiO_4$ in the range (0.12 TPa - 6300 K, 0.5 TPa-22000 K) and (0.2 TPa - 6300 K, 0.8 TPa-30000 K) respectively. This result is in disagreement with previous experimental data [*Spaulding et al.*, 2012] and recalls for a novel interpretation of the behavior of high pressure silicates liquids. In addition, no signature of the occurrence of MgO could be detected in both experimental data sets. Several hypotheses have been proposed to explain these observations, among which the stability of the undissociated $MgSiO_3$ and $Mg_2SiO_4$ liquids. Nevertheless, due to the importance of this debate further direct structural investigations under dynamic compression are mandatory.

Interestingly, for all the studied materials, it has been observed that metallization and melting do not occur at coincident thermodynamic conditions, implying the presence of poorly electrically conducting liquid in the phase diagram in the vicinity of the melting line. This result is of extreme relevance for the models of magnetic field generation via dynamo mechanism in a molten silicate layer, where a strongly enough conducting liquid is necessary to sustain the magnetic field. In addition, this behavior is different to what was established for $SiO_2$ [*Spaulding Thesis*; *Hicks et al.*, 2006 ], pointing out a difference in the electronic structure changes at melting between MgO, $MgSiO_3$ and $Mg_2SiO_4$ on one hand and $SiO_2$ on the other hand. This is an extremely intriguing input for condensed matter studies and it encourages a deeper investigation of the high-pressure regime of these materials with X-rays diagnostics.

Altogether, these results propose a different interpretation of the MgO, $MgSiO_3$ and $Mg_2SiO_4$ phase diagram at extreme conditions that need to be accounted for when modelling super Earths. Liquid potentially formed in the deep mantle on telluric exoplanets may not show structural transition and will be poor electrical conductors close to the melting line. The different pa-rameters presented in our study will be important inputs to model dynamic and internal structure of Super Earths.

**Acknowledgements**

This research was supported by the PlanetLab program of the Agence Nationale de la Recherche (ANR) grant No. ANR-12-BS04-0015-04. Discussions with V. Recoules have been very helpful for ab-initio calculations. We would like to thank Mélanie Escudier (INSP, Paris 6 university) for her careful and precious help during polishing and preparation of the starting materials. We would like also to thank Laurent Cormier (IMPMC, Paris 6 university) for the synthesis of $MgSiO_3$ glass in high temperature furnace.

Laser-shock experiments were conducted under the joint research project of the Institute of Laser Engineering, Osaka University. This work was supported in part by JSPS KAKENHI Grant Number 15K13609, JSPS core to core program on International Alliance for Material Science in Extreme States with High Power Laser and XFEL, and the X-ray Free Electron Laser Priority Strategy Program at Osaka University from the Ministry of Education, Culture, Sports, Science and Technology (MEXT). This experiment has been performed thanks to collaborations supported by GDRI N° 118 MECMA TPLA. We acknowledge the support of the COST Action MP1208 "developing the physics and the scientific community for inertial fusion.

The data used are listed in the references, supplements and Gekko and LULI repository.

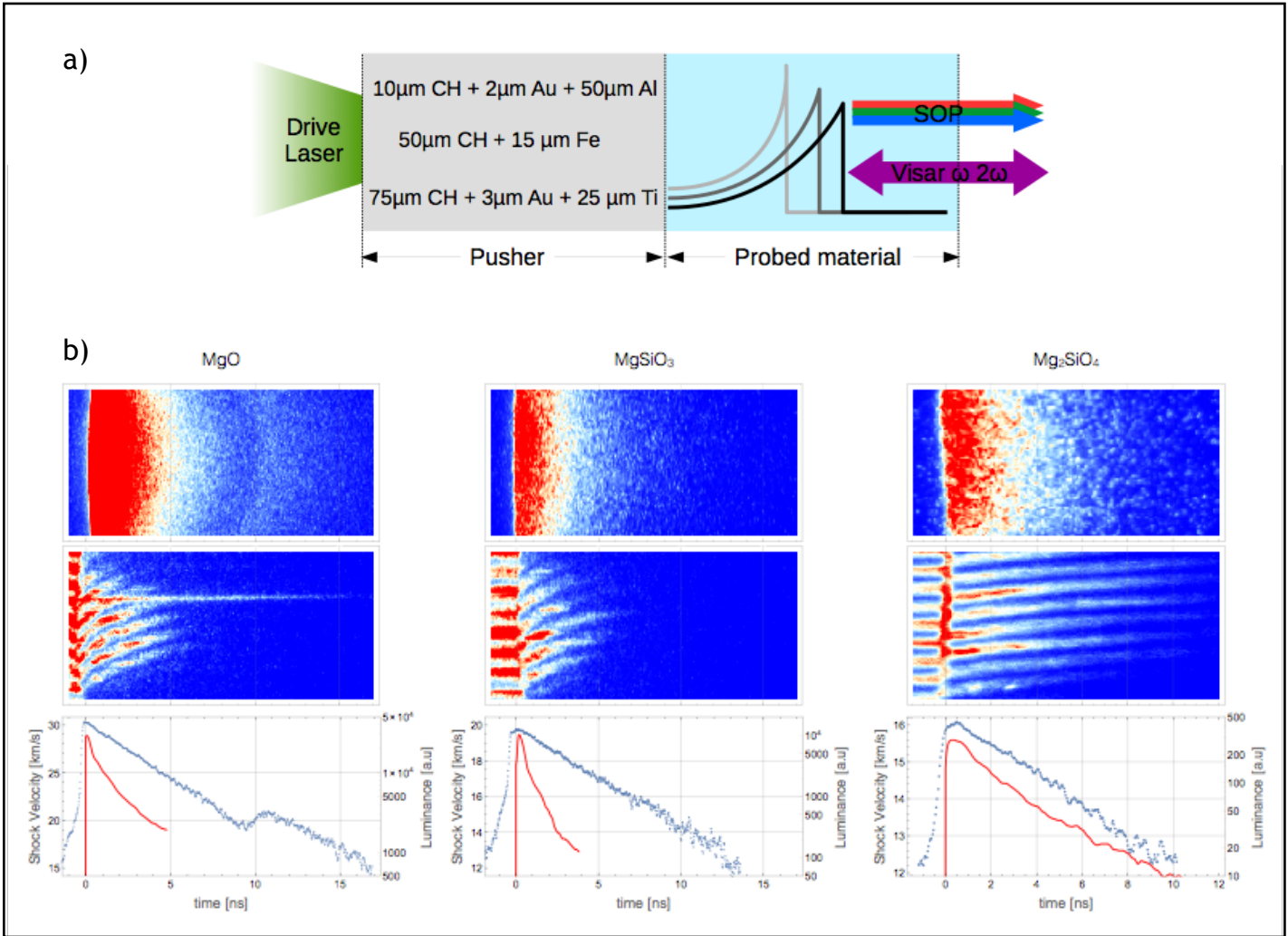

**Figure 1**. **(a)** Schematic of decaying shock experiments displaying the laser focused onto multi-layer target. The shock is generated into the pusher and then propagates into the material (MgO, MgSiO$_3$ or Mg$_2$SiO$_4$). Shock velocity and reflectivity are measured by VISARs at 532 nm and 1064 nm, while thermal emission is measured by an SOP. **(b)** Typical SOP and VISAR signals are shown with thermal emission (blue dots) and shock velocity (red line) time profiles for each materials. The represented shots were performed with a 1.2 ns pulse with 726.7 J and 236 J for MgO and MgSiO$_3$ respectively, and with a 2.5 ns pulse with 990 J Mg$_2$SiO$_4$.

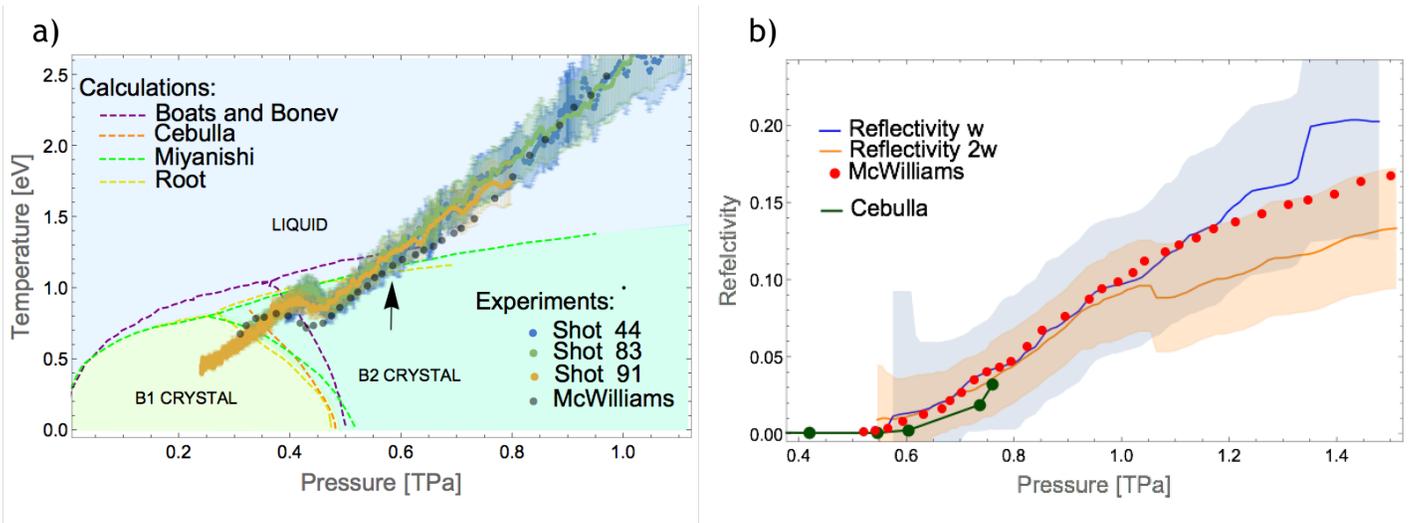

**Figure 2**. **(a)** The experimental P-T curves measured along the Hugoniot at LULI (coloured dots) are represented in the MgO P-T diagram in comparison with previous experiment (black dots) [*McWilliams et al.*, 2102] and ab-initio calculations (coloured dashed lines) [*Boates and Bonev*, 2013; *Cebulla and Redmer*, 2014; *Miyanishi et al.*, 2015; *Root et al.*; 2015]. **(b)** MgO reflectivity at 532 nm and at 1064 nm are represented in orange and blue respectively, in comparison with previous experimental results (red dots) [*McWilliams et al.*, 2012] and calculations (green joined dots) [*Cebulla and Redmer*, 2014].

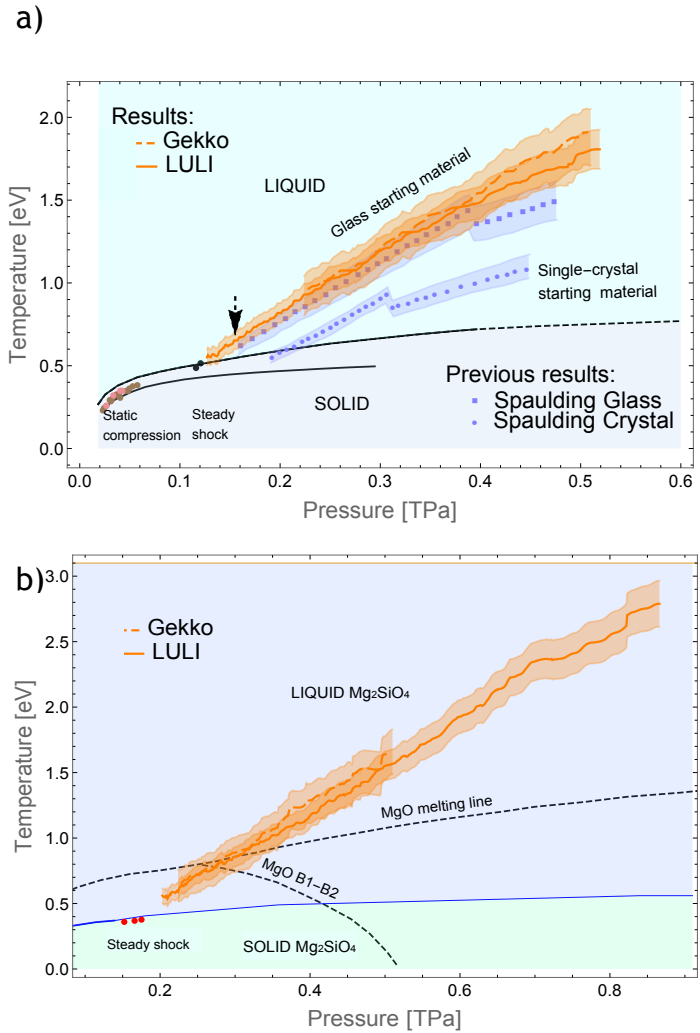

**Figure 3**. **(a)** The experimental MgSiO$_3$ glass P-T curves measured along the Hugoniots at LULI (orange full line) and at Gekko (dotted orange line) are represented in MgSiO$_3$ phase diagram in comparison with previous experimental results (blue disk and square) [*Spaulding et al.*, 2012]. The arrow indicates the beginning of the reflecting segment along the MgSiO$_3$ Hugoniot. **(b)** The experimental Mg$_2$SiO$_4$ hugoniots measured at LULI (orange full line) and at Gekko (dotted orange line) are represented in the Mg$_2$SiO$_4$ phase diagram. Dashed lines show MgO coexistence lines.